\documentclass[preprint,12pt]{elsarticle}

\usepackage{graphicx}

\usepackage{amssymb}
\usepackage{amsthm}
\usepackage{amsmath}

\usepackage{pdfpages,multirow,ragged2e} %
\usepackage{adjustbox}
\usepackage{placeins}

\journal{Computer Physics Communications}

\begin{document}

\begin{frontmatter}



\title{Beam Dynamics module of RFQ design software Demirci}

\author[Emre1,Emre2]{E. Celebi}
\ead{emre.celebi@cern.ch}
\author[Gorkem]{G. Turemen}
\author[Orhan]{O. Cakir}
\author[Gokhan]{G. Unel}

\address[Emre1]{Bogazici University,Department of Physics, Istanbul, Turkey}
\address[Emre2]{Istanbul Bilgi University, Energy Systems Engineering Department,Istanbul, Turkey}
\address[Gorkem]{Turkish Energy, Nuclear and Mineral Research Agency, Ankara, Turkey}
\address[Orhan]{Ankara University, Department of Physics,
Ankara, Turkey}
\address[Gokhan]{University of California at Irvine,Department of
Physics and Astronomy, Irvine, USA}

\begin{abstract}
\emph{The RFQ design tool DEMIRCI aims to provide fast and accurate
simulation of a light ion accelerating cavity and of the ion beam
in it. It is a modern tool with a graphical user interface leading
to a ``point and click'' method to help the designer. This article
summarises the recent software developments such as the addition
of RFQ acceptance match, beam dynamics and 8-term potential coefficient
calculations. Its results are compared to other similar software, to discuss the attained compatibility level. }
\end{abstract}
\end{frontmatter}
\section{Introduction }

The acceleration of ion beams has become more efficient with the utilisation
of the radio frequency quadrupole (RFQ) cavities since 1970s. The
theoretical work on the RFQ cavity which bunches, focuses and accelerates
was first performed by Kapchinsky and Tepilyakov \cite{TepilKap}.
The solutions of the Poisson equation defining electric
potential inside the cavity is still commonly referred to as the K-T
potential to honor them. 
The K-T potential being an infinite series, its first two
terms are generally used in initial simpler designs in two dimensions;
6 more terms are added for taking into account the geometrical effects
in three dimensions and finally more terms could be added to take
into account the production and assembly deficiencies which would
break the four fold axial symmetry of the cavity. Therefore, the successful
operation of an RFQ relies on the correct determination of the K-T potential
terms.

With the ever increasing requirements from the community, 
the need for  software tools that would yield an 
easy and accurate RFQ design in a rapid and user friendly manner 
is ever-growing. Currently, there are a limited number
of such software tools, some written almost 30 years ago,
with PARMTEQM suite and LIDOS being the leading examples \cite{Parmteq,Lidos}.
Moreover, each such tool is specialising on a particular aspect of
the design process, e.g. TOUTATIS on the beam dynamics and error studies.  
Using different simulations at each design stage
and manually keeping track of these software tools is error prone and time consuming. 
DEMIRCI RFQ designer software was born out of such a need 
for a simple-to-use tool with multiple capabilities \cite{demirpaper}. 
In its first version, running on
Unix-like systems, it was possible, even for the most inexperienced 
user, to prepare a simple design based on the two term potential using
the graphical means provided by the ROOT libraries from CERN \cite{root}.
The next version contained enhancements like the enlargement of the
variable set that can be plotted, the multi-lingual graphical user
interface and instruction manuals and the availability in Linux and
Windows operating systems \cite{Dem-recent}. 
In this manuscript, the recently developed DEMIRCI PRO's beam dynamics module
is presented. It contains additions such as eight term potential
multipole coefficients calculation, beam dynamics simulations with
eight or two terms and RFQ acceptance calculations.

\section{PRO version of Demirci}

Recently DEMIRCI has been updated with the calculation of the eight
term (8T) K-T potential multipole coefficients, beam dynamics calculations, ion beam visualisations using the 8T potential and RFQ acceptance and mismatch factor calculations. The validity of these calculations was tested by comparing its results to other similar software. The speed and accuracy of the calculation technique is also investigated as an optimization process. The remainder of this section discusses these additions.

\subsection{Eight Term Potential}

To realistically design an RFQ which typically consists of about hundred
cells, one needs to correctly and rapidly determine the time independent
part of the electrical K-T potential for each cell : 

\begin{eqnarray}
U(r,\theta,z)= \frac{V}{2}[\sum_{m=1}^{\infty}A_{0m}(\frac{r}{r_{0}})^{2m}\cos(2m\theta)\label{eq:generic-potential}\\
\qquad + \sum_{n=1}^{\infty}\sum_{m=0}^{\infty}A_{nm}I_{2m}(nkr)\cos(2m\theta)\cos(nkz)]\nonumber 
\end{eqnarray}

where $r$ and $\theta$ are cylindrical coordinates for which $z$
represents the beam direction, $V$ is the inter-vane voltage, $k$
is the wave parameter given by $k\equiv2\pi/\lambda\beta$, with $\lambda$
being the RF wavelength and $\beta$ being the speed of the ion relative
to the speed of light. Also, $r_{0}$ is the mean aperture of the
vanes, $I_{2m}$ is the modified Bessel function of order $2m$ and
the $A_{nm}$ are the multipole coefficients whose values, depending
on the vane geometry, should be obtained. 
Only terms that are reflecting the cell geometry symmetry are considered for modelling the cells. Moreover, although the summation has infinite number of terms, only first 8 coefficients are used to model electric potential inside the cell. Those coefficients are ;
A01, A10, A10, A11, A12, A21, A23, A30 and A32.\cite{ParmteqManual}
Legacy software like PARMTEQM \cite{Parmteq} relies on
data from pre-computed tables containing image charge calculations
using the integral method to determine the multipole coefficients.
Another method is to use a 3 dimensional differential finite element
method (FEM) \cite{FB} to obtain the potential distribution across the RFQ length
and then to find the multipole coefficients via a least squares
fit to FEM nodes \cite{FEM-first}. Since this method is reported (in \cite{FEM-first}) 
to be as accurate as the image charge method and to be much faster from a
computational point of view, an independent implementation of this approach
is performed in DEMIRCI libraries.

\begin{figure}
\begin{centering}
\includegraphics[width=0.5\columnwidth]{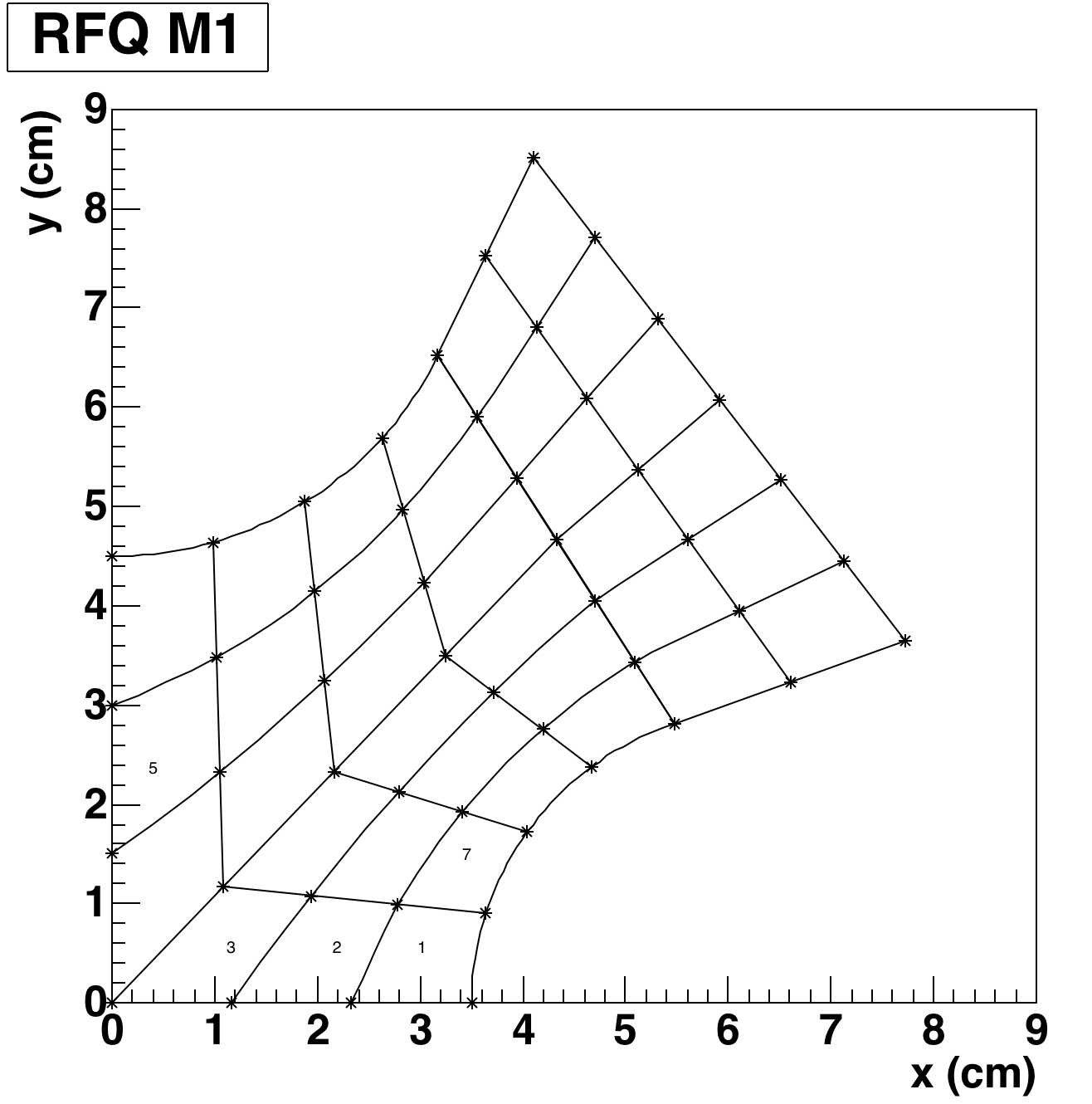}\includegraphics[width=0.5\columnwidth]{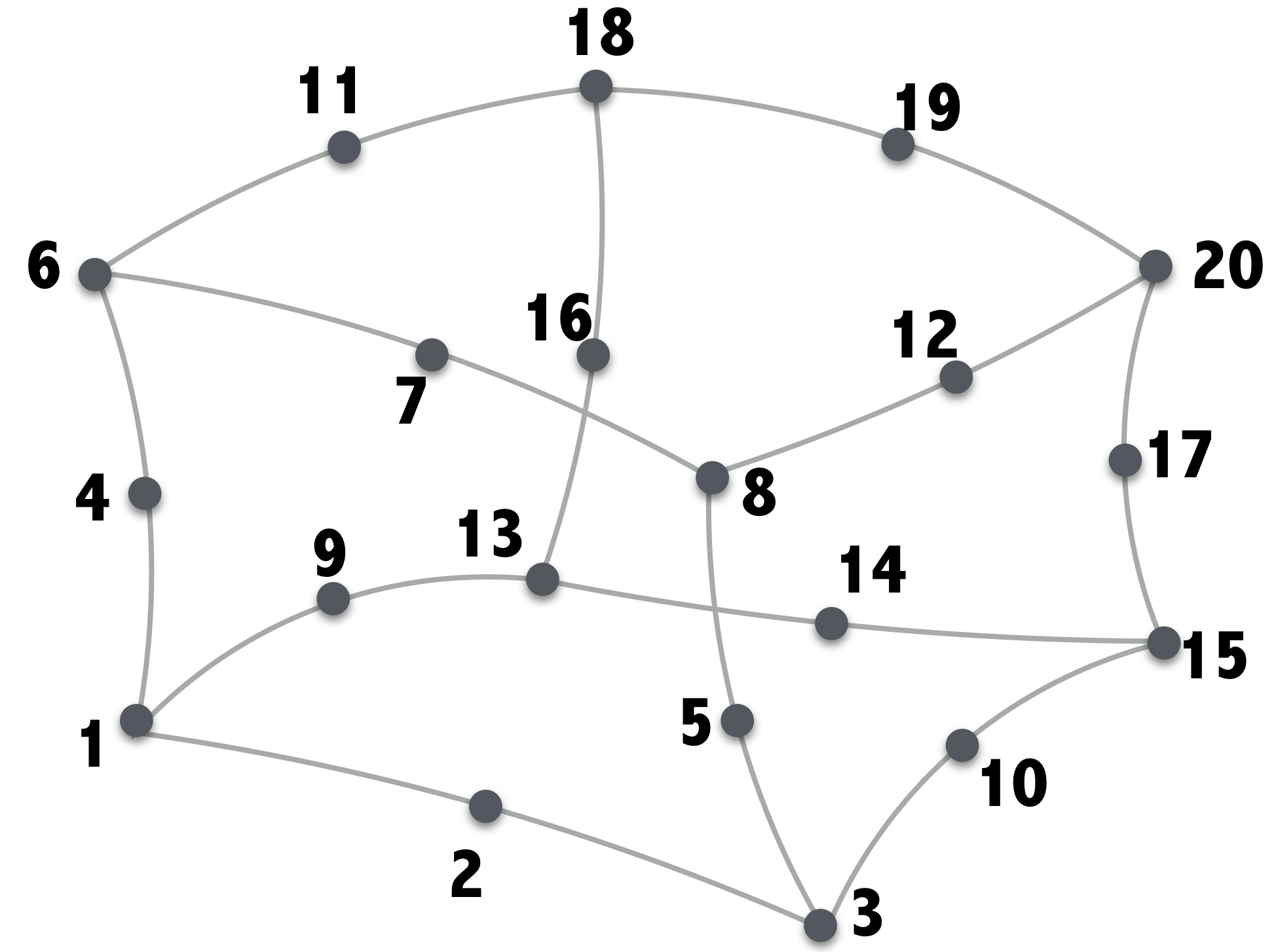}
\par\end{centering}
\caption{The meshing in 2D on the left showing the one quarter of the RFQ cross
section focusing on vane tips and on the right the 3D meshing unit
with a 20 node isoparametric solid brick element. \label{fig:meshing}}
\end{figure}

The FEM technique requires a 3-dimensional meshing of the volume at
which the Poisson equation is to be solved. However, the work starts with
dividing the two dimensional the cross section (quarter view due to symmetry) of the vane tips into equal length segments as seen in Figure\ref{fig:meshing} left side. As seen, the Cartesian coordinates are effectively transformed into curved coordinates $u,v$ which follow pole tip surface and its normal. Here, all the segments including the intra-pole ones, are constructed by interpolation from the pole contours. 

A similar division is also made along the RFQ axis, yielding 20 node isoparametric
bricks as the mesh units as shown in the same figure, right side. The nodes are defined as the line end and middle points.
The number of meshing units define both the accuracy of the calculations
and the calculation speed. Although the details will be presented
in the next section, a good compromise value has been determined and set  in the code as the default number
of 3-D mesh units: $10\times10\times10$ 
in $x$, $y$ and $z$ coordinates. Once the
meshing is complete, the general stiffness matrix is prepared from
individual isoparametric bricks. For each mesh unit, one starts by defining an associated Cartesian cube in natural coordinates which can be transformed into an isoparametric brick via an interpolation function.
The local stiffness matrix is defined via the integration of these interpolation functions in natural coordinates. 
Once the local matrices are defined, they are stitched together into a general stiffness matrix so that nodes with multiple cell memberships are expressed only once. The indexing for all nodes follows a specific algorithm. It starts with the first node located at (0,0,0) where these three digits represent the curvilinear coordinates $u$, $v$ and $z$. Other nodes are added by first looping over $u$, then $v$ and finally $z$. Since node indexing follows this predefined algorithm,
the index of any node can be easily calculated for a given geometrical position. By virtue of this property, nodes that lie on the boundary surface (namely the vane tips) can be determined easily and  their values can be set using the boundary conditions on the pole tips. 

After these steps, a general stiffness matrix $A$ is obtained. It relates node values to the solution vector ($x$) and to electric charge inside the volume ($b$): $Ax=b$. Since initially there are no particles inside the RFQ volume, one needs to work with the Poisson equation, therefore all elements of the vector $b$ are equal to zero. 
Since the solution vector contains both unknown and known values (due to boundary conditions), we would like to reduce the previous equation to $ A'x'=b'$ form, where  known elements of $x$ (and their counterparts in $A$) are removed and the new indexing is mapped to old indexing as a separate dictionary for further use. 
The new vector $b'$ is calculated by using the multiplication of known $x$ values with their associated elements of the general stiffness matrix $A$.
\begin{table}
\caption{Definitions of the RFQ cells used for comparison. Minimum bore radius and cell length are in cm.\label{tab:Ref-Cells}}
\centering{}%
\begin{tabular}{|c|c|c|c|}
\hline 
cell\# & $a$ & $m$ & $\ell_{cell}$\tabularnewline
\hline 
\hline 
20  & 0.409 & 1.020 & 0.58\tabularnewline  \hline 
60  & 0.399 & 1.072 & 0.60\tabularnewline \hline 
100 & 0.392 & 1.111 & 0.68\tabularnewline \hline 
140 & 0.381 & 1.171 & 0.92\tabularnewline \hline 
180 & 0.309 & 1.631 & 1.94\tabularnewline \hline 
\end{tabular}
\end{table}
The resulting Poisson equation is solved numerically for each node to find the electrostatic potential in each cell, using the conjugate gradient technique. 
Conjugate gradient method is widely used to solve this type of problems since it does not need any matrix inversion, a technique that changes the matrix sparseness by creating new nonzero elements. 
Depending on the matrix size, the inversion can be problematic as storing intermediate values quickly becomes impractical even on the disk, let alone in memory. 
On the other hand, the conjugate gradient method is an iterative approach. It starts by guessing an initial solution ($x_g'$) and calculates its associated solution vector ($b_g$) to compare it to the known solution vector $(b')$.
The algorithm terminates when the current guessed solution ($b_g$) is close enough to original solution vector ($b'$).
The solutions for the nodes within the minimum bore radius are
then fitted to the 8T potential function with the least squares method
to find the first 8 K-T potential multipole coefficients. 

To test the validity of the procedure and the correctness of the implementation,
the multipole coefficients calculated by DEMIRCI are compared to the
ones calculated by other groups as found in the literature \cite{FEM-first}.
Five reference cells for which the coefficients are calculated range
from low $m$ to high $m$, i.e. from the entrance of the RFQ
to its exit. The properties of these reference cells are given in
Table \ref{tab:Ref-Cells}.

The multipole coefficient calculation results for the five reference cells
are shown in Table \ref{tab:Comparison-table}. Each block of four rows
contains the calculation results of the multipole coefficients of
the reference cell indicated by the first column.
The publication in \cite{FEM-first} compares the coefficients obtained using RFQCoef program which in turn uses a 3D FEM technique, to those calculated by the CHRG3D (from LANL) program which uses the image charge method.
The present note adds two more coefficient calculations to that set. The first set comes from PARMTEQM, a modernized version of the CHRG3D and the calculations using DEMIRCI. In order to decide on the quality of calculations, some  comparison criterion has to be defined. 
We defined a total relative error as 
$\epsilon_{tot}=\sqrt{\underset{i}{\sum}(\frac{C_{i}-C_{i}^{R}}{C_{i}^{R}})^{2}}$
where the summation runs over the first 8 coefficients and  $C^{R}$ represent the reference values. In the case where the reference value is zero, the division becomes undefined. Although there are multiple propositions to handle this scenario, in this note the simplest alternative namely the test value ($C_i$) is used in denominator.
 By simply comparing the total error in the last column one is able to judge on the quality of the 8T potential calculations. To that end, results from all programs are compared to CHRG3D results, additionally DEMIRCI results are also compared to PARMTEQM for which the total relative error is given in parentheses.
One can immediately notice that the newer version of LANL tool, PARMTEQM, is always more compatible with the CHRG3D as compared to RFQCoef except the last cell which will be discussed later. This is expected since both CHRG3D and PARMTEQM originate from the same scientific collaboration. RFQCoef results are in good agreement with CHRG3D in all 5 test cells.
Except for the last cell (180), DEMIRCI results are compatible with both older and newer versions of the LANL software, the agreement being better with PARMTEQM. Moreover, except the last cell it always reports an error equal to or smaller then the RFQCoef results. The last cell seems to be peculiar in the sense that both PARMTEQM and DEMIRCI report unprecedentedly large total errors while these two being in agreement with each other. As mentioned before, for this cell, CHRG3D and RFQCoef are also in agreement with each other. Since we do not have the exact calculations of the publication \cite{FEM-first}, we can only speculate about this strange behaviour. The most reasonable explanation is that the older software (CHRG3D and RFQCoef) shared a common calculation error which was subsequently corrected. Another, less likely, possibility is a typo in \cite{FEM-first} in the definition of cell 180. Independent of this situation, the analysis shows that DEMIRCI is able to calculate the 8 term coefficients in good accuracy with respect to the industry standard software from LANL.

\begin{table*}
\caption{Comparison of the multipole coefficients calculated by available software. The errors are always calculated relative to CHRG3D except for DEMIRCI which also reports relative to PARMTEQM in parentheses.  
\label{tab:Comparison-table}}
\centering{}%
\begin{adjustbox}{width=1\textwidth}  
\small
\begin{tabular}{|c|l||c|c|c|c|c|c|c|c||c|}
\cline{1-1} \cline{3-11} 
\# & \multicolumn{1}{c|}{} & $C_{10}$  & $C_{00}$/$a^{2}$  & $C_{11}$  & $C_{01}/a^{6}$  & $C_{30}$  & $C_{20}$  & $C_{31}$  & $C_{21}$  & $\epsilon_{tot}$\tabularnewline
\cline{1-1} \cline{3-11} 
\multicolumn{1}{c}{} & \multicolumn{1}{c|}{} & $A_{10}$ & $A_{01}$ & $A_{12}$ & $A_{03}$ & $A_{30}$ & $A_{21}$ & $A_{32}$ & $A_{23}$ & \tabularnewline
\hline 
 & CHRG3D & 0.00606 & 5.73007 & 0.05304 & 4.94852 & 0.0 & -0.00003 & 0.0 & 0.00072 & -\tabularnewline
\cline{2-11} 
 & RFQCoef  & 0.00601 & 5.74846 & 0.05031 & 4.41814 & 0.0 & -0.00002 & -0.00001 & -0.00077 & 2.32\tabularnewline
\cline{2-11} 
20 & ParmteqM & 0.00539 & 5.72978 & 0.06728 & 4.9641 & 0.0 & 0.00001 & 0.0000 & 0.00069 & 1.36\tabularnewline
\cline{2-11} 
 & Demirci   &0.00549&	5.73015&	0.06511&	4.92675&	0.00000&	0.00&	-0.00000&	-0.00000 & 1.43 (1.41)\tabularnewline
 
\hline 
\hline 
 & CHRG3D & 0.02273 & 5.73981 & 0.20874 & 4.89934 & 0.00000 & 0.00003 & 0.00000 & -0.00031 & -\tabularnewline
\cline{2-11} 
& RFQCoef  & 0.02280 & 5.75573 & 0.21751 & 4.51230 & 0.00000 & -0.00006 & -0.00001 & 0.00402 & 14.32\tabularnewline
\cline{2-11}
60 & ParmteqM  &0.02051&	5.73962&	0.27159&	4.98538&	0.00000&	0.00007&	0.00000&	0.00134 & 5.50\tabularnewline
\cline{2-11} 
 &  Demirci   &0.0206495&	5.74013&	0.26736&	4.98026&	0.00000&	0.00005&	0.00001&	0.00091& 4.13 (1.08)\tabularnewline
\cline{2-11} 
\hline 
\hline 
 & CHRG3D & 0.04307 & 5.74320 & 0.45675 & 4.88818 & 0.0 & 0.00018 & 0.0 & 0.00087 & -\tabularnewline
\cline{2-11} 
& RFQCoef & 0.04296 & 5.85630 & 0.48357 & 4.03974 & -0.00001 & 0.00022 & -0.00002  & 0.00628 & 6.38\tabularnewline
\cline{2-11}
100 & ParmteqM  &0.03866&	5.74426&	0.63987&	4.99520&	0.00000&	0.00028	&0.00000&	0.00443 & 4.15\tabularnewline
\cline{2-11} 
 &  Demirci   & 0.03896&	5.74496&	0.63007&	5.01070&	0.00000&	0.00027&	-0.00001&	0.00262 & 2.33 (1.08)\tabularnewline
\hline 
\hline
 & CHRG3D & 0.09684 & 5.74616 & 1.40462 & 4.74544 & 0.0000 & 0.00076 & 0.0000 & -0.04081 & -\tabularnewline
\cline{2-11} 
& RFQCoef  & 0.09662  & 5.75831 & 1.35976 & 3.84250 & 0.00000  & 0.00096 & -0.00015  & -0.09994 & 1.79\tabularnewline
\cline{2-11} 
140 & ParmteqM  &0.08634&	5.75451&	2.64614&	4.95261&	0.00001&	0.00165&	0.00004&	0.01875 & 2.51\tabularnewline
\cline{2-11} 
 &  Demirci   & 0.08687&	5.75461&	2.60377&	4.95999	&0.00001	&0.00163	&0.00004&	0.01968 &2.50 (0.51) \tabularnewline
 \cline{2-11} 
\hline 
\hline
 & CHRG3D & 0.44918 & 5.65552 & -36.14094 & 2.36514 & -0.00002 & -0.04355 & -0.10643 & -100.715 & -\tabularnewline
\cline{2-11} 
& RFQCoef  & 0.44807 & 5.66045 & -34.26460 & 1.44060 & -0.00007 & -0.04664 & -0.10646 & -99.0374 & 2.53\tabularnewline
\cline{2-11}
180 & ParmteqM  &0.39947	&5.80745&	81.02745&	3.70216	&0.00078	&0.05586	&-0.03312&	-49.50706& 40.21\tabularnewline
 \cline{2-11} 
 & Demirci  & 0.40167&	5.79689&	75.01900&	3.73725&	0.00071&	0.05036&	-0.03615&	-38.81840& 36.64 (0.281)\tabularnewline
\hline 
\end{tabular}
\end{adjustbox}
\end{table*}

\begin{figure}[!htb]
\centering \includegraphics[clip,width=1\columnwidth]{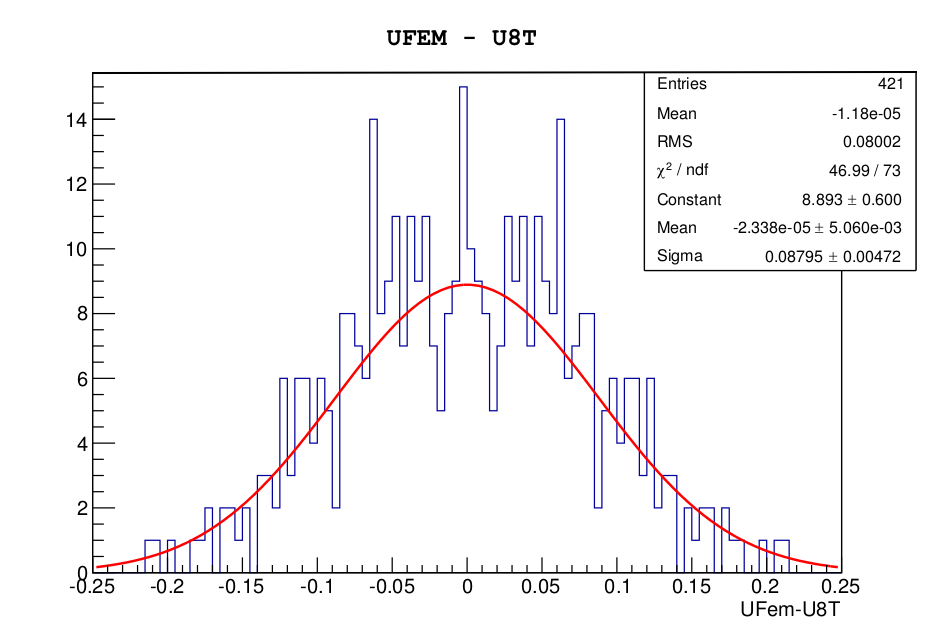}
\caption{The cell number finite element method minus reported 8T potential
for cell 20 from Table \ref{tab:Ref-Cells} }
\label{hucre20} 
\end{figure}

As an additional check, the electric potentials obtained from the
FEM solutions using CST Studio Suite\textsuperscript{\textregistered} \cite{CST} on all nodes are compared to fit results for a random
cell. The histogram of the difference between the two are shown in
Figure \ref{hucre20} together with a Gaussian function fit. 
The distribution shows that the error originating from the fitting
procedure is about 1/1000. Even if in some long cells, the 8T potential
is not adequate for defining the equipotential surface, the error
is smoothed out by the fitting procedure and its order of magnitude
is expected to remain at this level.

\subsubsection{Calculation performance}
\begin{figure}[!b]
\centering \includegraphics[clip,width=1\columnwidth]{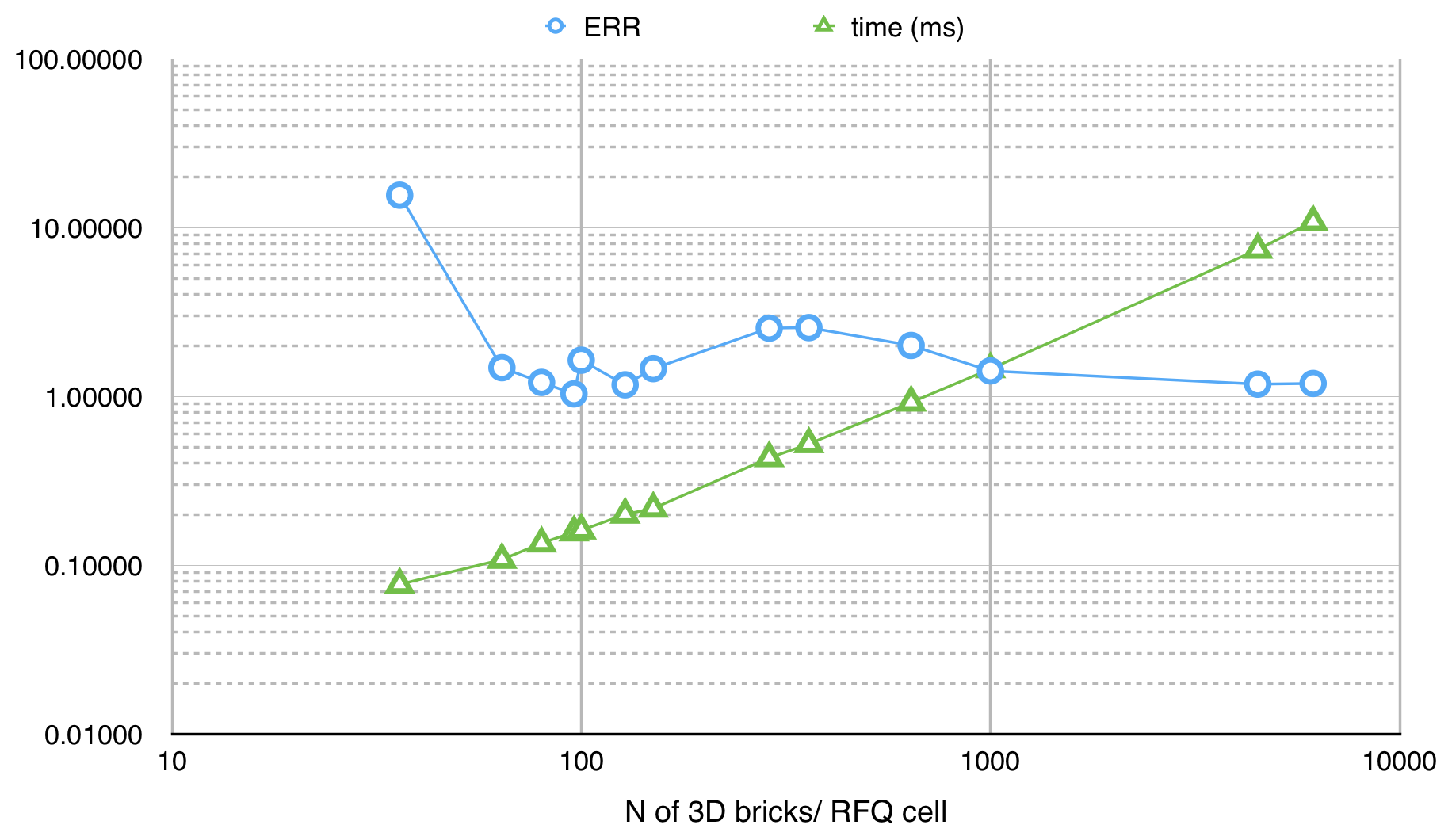}
\caption{The computing time in $ms$ and the associated error in 8T potential
multipole calculation for cell 100 as a function of the FEM meshing granularity.}
\label{hucre100-timing} 
\end{figure}
The calculation performance is estimated by varying the mesh size
for the test cell (number 100) from $3\times3\times4$ up to $16\times16\times24$
where the numbers are for the number of divisions in $x,y$ and $z$
directions respectively. The calculation time is defined as the time
reported by the operating system (OSX 10.15) on an Intel i7 3 GHz laptop computer, as the
sum of the time spent in the user and kernel spaces. The results are
shown in Figure~\ref{hucre100-timing}. The horizontal axis is the direct
multiplication of the scanned division values in each coordinate,
therefore represents the number of 20-node isometric bricks ($N$)
used to compute the electric potential in cell 100. One easy observation
is the linear increase in the computation time as shown by the green
triangles. Therefore it is important to get the best accuracy with
the smallest number of bricks. For an accuracy estimation, in each
case, an associated error is also calculated with respect to CHRG3D results as $\epsilon=0.1\times\epsilon_{tot}$ . The coefficient
0.1 is employed only to scale the error values, shown by blue circles,
such that these can be plotted into the same $y$ axis range. Since
the problem has $x-y$ symmetry, these two axes were always split
to the same number of divisions, however the number of $z$ axis divisions
was scanned independently. The smallest error, as obtained with a very
large number of (O(6000)) bricks, is also attained with just below 100 bricks which
can be defined as the optimum operating point. The worsening of the
accuracy for the region $N$\textasciitilde{}100 is related to the
ratio $N_{z}/N_{x}$.

\subsection{Beam Dynamics}

One of the DEMIRCI's features is the possibility to insert a number of macro
particles generated randomly according to the beam phase space definitions
just at the entrance of the RFQ. The number of the particles, their
phase space properties (beam emittance and Twiss parameters) and the beam
shape (Gauss, flat, water-bag) are all user selectable variables.
The purpose of these beam dynamics simulations is to make the overall
RFQ design process more interactive by monitoring the ion beam behaviour
in phase and normal spaces along the RFQ. After the initialisation
step, DEMIRCI employs a time based approach, in contrast to the position
based one in PARMTEQM, to track the individual macro particles. Each
such particle's position, velocity and energy is calculated at each
time ($t$) to find the equivalent quantities after a small enough
time interval ($\delta t$). DEMIRCI has been written to allow the
user for selecting the electric field ($\vec{E}$) values obtained
from two term or eight term potentials. Therefore the relevant equations
for position, velocity and kinetic energy at the next time interval
($t+\delta t$) are :

\begin{eqnarray}
\vec{x}^{t+\delta t} & = & \vec{x}^{t}+\vec{\beta}^{t}c\delta t+{\frac{1}{2}}{\frac{Q}{\gamma^{t}M}}\vec{E}^{t}\delta t^{2}\quad,\\
{\gamma}^{t+\delta t}\vec{\beta}^{t+\delta t} & = & \gamma^{t}\vec{\beta}^{t}+{\frac{1}{2}}{\frac{Q}{cM}}\delta t(\vec{E}^{t}+\vec{E}^{t+\delta t})\quad,\\
E_{k}^{t+\delta t} & = & (\gamma^{t+\delta t}-1)Mc^{2}\quad,
\end{eqnarray}

\begin{figure}[!b]
\centering \includegraphics[clip,width=80mm]{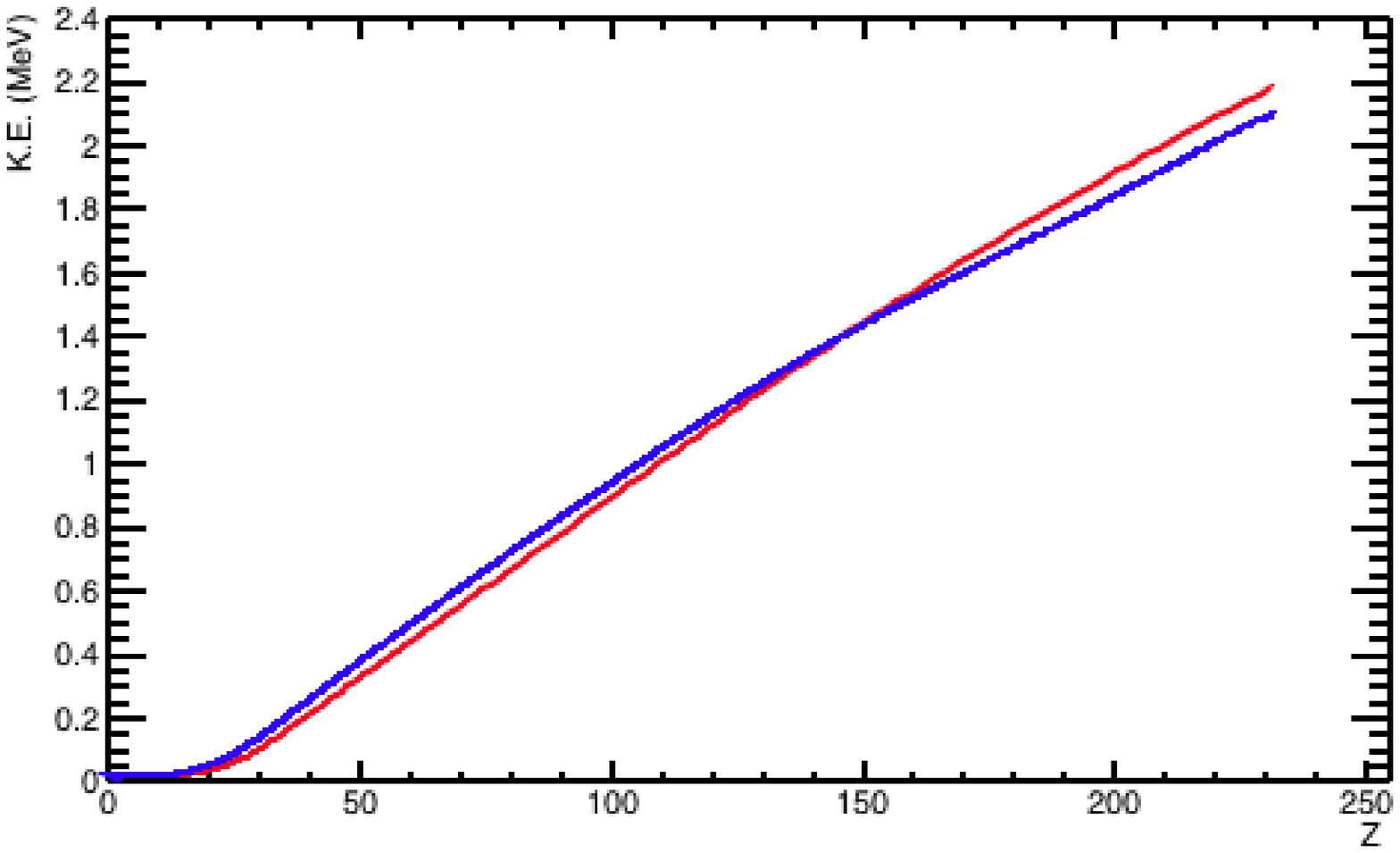}

\includegraphics[clip,width=80mm]{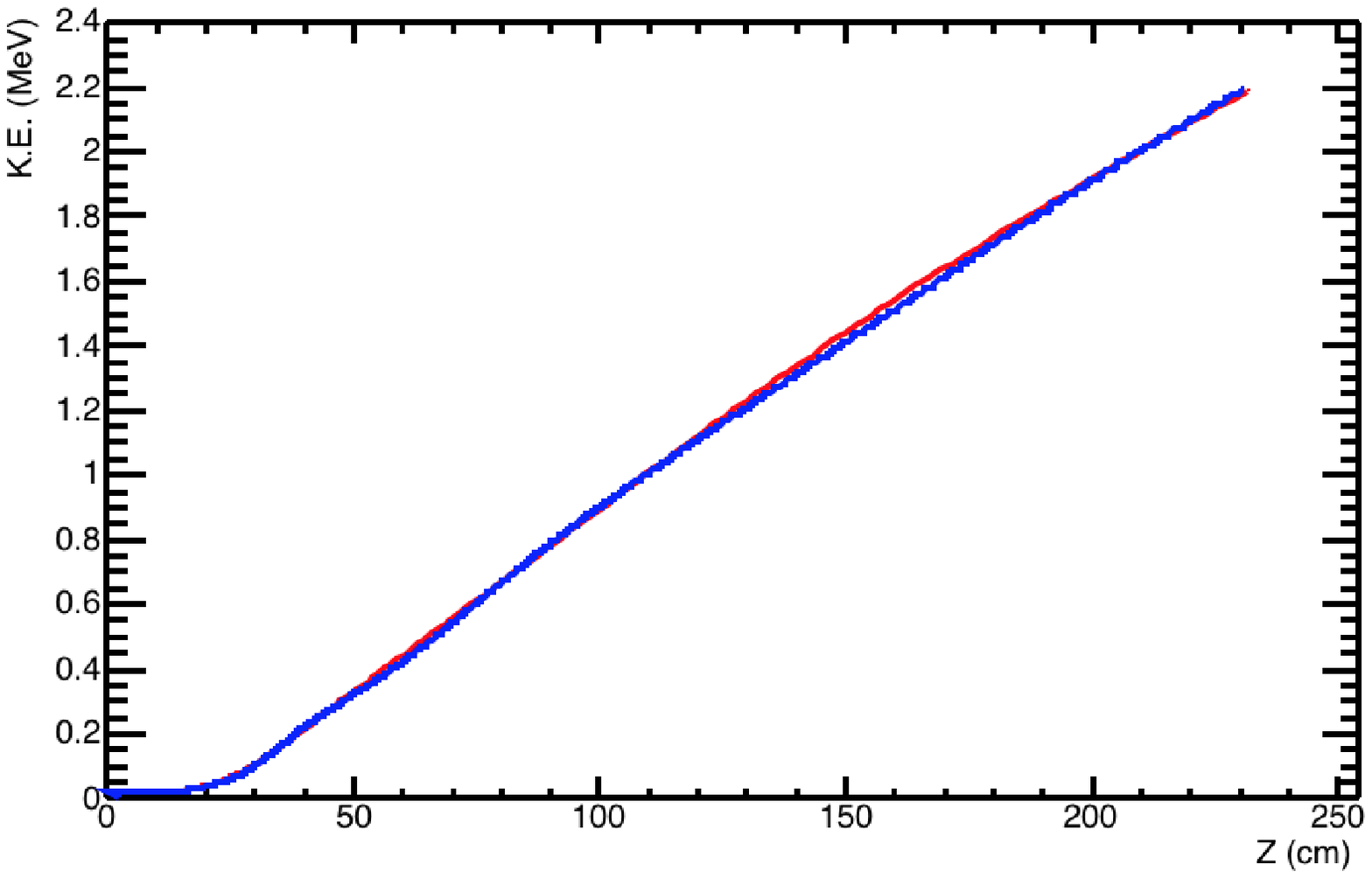}

\caption{Average kinetic energy (in blue) of the macro particles obtained from
dynamic equations in blue compared to estimations (in red) from averaged
design equations. Top plot is with 2T potential whereas the bottom
one is with 8T one.}
\label{KE} 
\end{figure}

where quantities with superscript $t$ represent their values at time
$t$, whereas $Q$ and $M$ are the charge and mass of the macro-particle
and $c$ is the speed of light. The DEMIRCI library has been enlarged
with functions that calculate and plot the particle trajectories according
to the above equations. Although this functionality has been previously
reported\cite{Dem-recent}, it is revisited in this note, since the
user can now select between the 2T potential and 8T potential.

\begin{figure}[!htb]
\begin{centering}
\includegraphics[width=0.95\columnwidth]{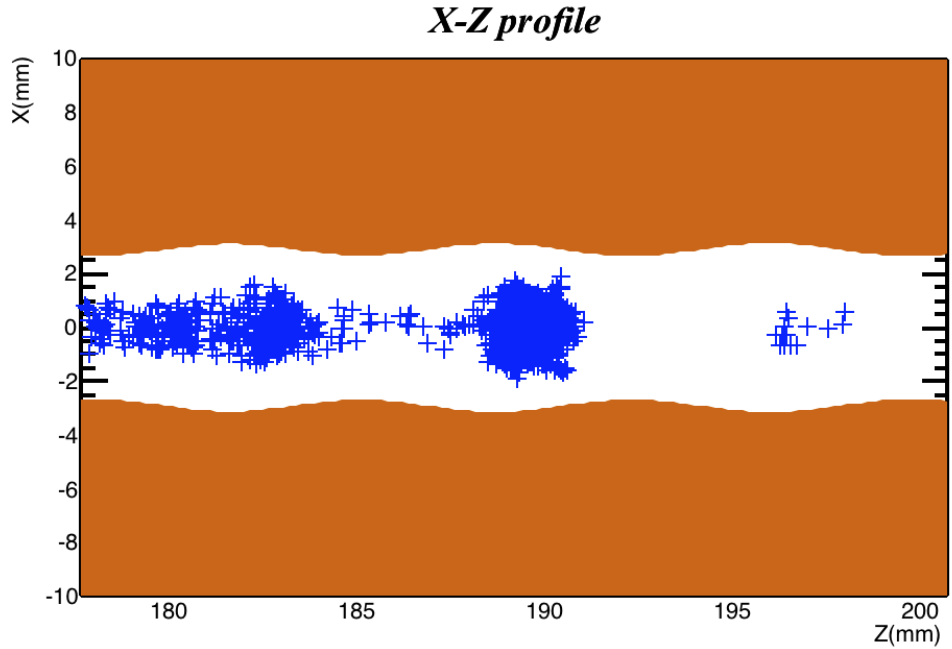}
\par\end{centering}
\caption{bunching of particles in an RFQ becomes visible. }
\label{fig:bunch} 
\end{figure}

Three beam dynamics panels, selectable from the main DEMIRCI window,
allow monitoring the macro-particle behaviour along the RFQ using
the equation set given above.

\subsubsection{Comparisons with similar software}

These panels become usable only after
an RFQ description is loaded and its overall behaviour is calculated,
including the multipole coefficients of the 8T potential. Each beam
dynamics panel contains four sections showing any of the predefined
distributions selectable by the designer. The currently available
distributions are: spatial and angular distributions in both vertical
and horizontal planes, beam profiles and emittances in all 3 directions,
and the average kinetic energy. Figure~\ref{KE}
contains two such plots for an example design at 352.2 MHz:
the test particles' average kinetic energy from
the static equations in red can be compared to the value, in blue,
obtained from dynamic calculations on the top plot using 2T potential
and on the bottom using the 8T potential. 
The bunching of the particles as they move along the RFQ can also be
observed. Figure~\ref{fig:bunch} shows the behaviour of the macro-particles
in the $x-z$ plane, using electric fields from the 8T potential calculations.  
These simulations can be started,
stopped and reinitialised by the user at any time. The default time
step for calculations and visualisation is 0.1~ns. 
However, displaying the individual behaviour of a large number of particles 
could be more than the average CPU or graphics processor can handle. 
To overcome this problem, the plot refresh rate is made 
user selectable using a scroll bar on screen
such that adjusting it would allow the user to play the simulation
rapidly (up to 10 times) and to slow it down when needed.\\

\begin{figure}[!t]
\begin{centering}
\includegraphics[clip,width=80mm]{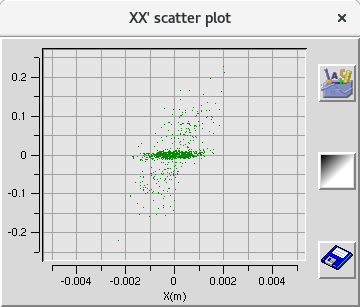}
\par\end{centering}
\begin{centering}
\includegraphics[clip,width=80mm]{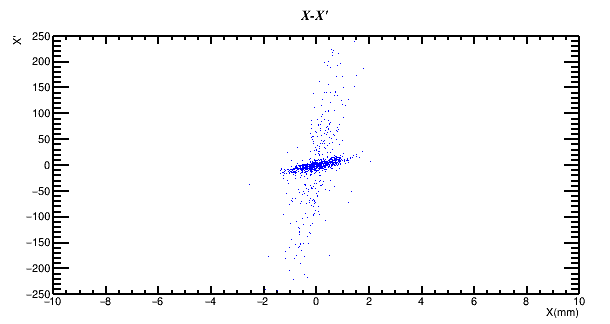}
\par\end{centering}
\caption{$x-x'$ phase space comparison between DEMIRCI and TOUTATIS \label{fig:xx-phase}}
\end{figure}

As a further test of the multipole coefficient calculation technique's
implementation, the same RFQ design has been studied both with DEMIRCI
and TOUTATIS \cite{toutatis}. Figure \ref{fig:xx-phase} shows the $x'x$ phase space
at around $z$=2300 mm as obtained from both software. For further
reference, $x'$ is defined as \cite{linacbook}:

\begin{equation}
x'=(x_{f}-x_{i})/(z_{f}-z_{i})
\end{equation}

where the subscript $f$ ($i$) stands for the next (current) position
of a particle. 

\begin{figure}[!t]
\begin{centering}
\includegraphics[clip,width=80mm]{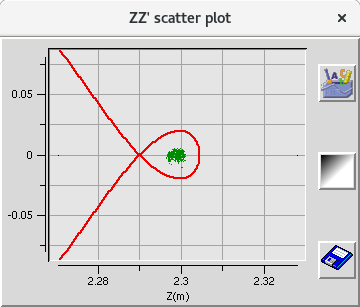}
\par\end{centering}
\begin{centering}
\includegraphics[clip,width=87mm]{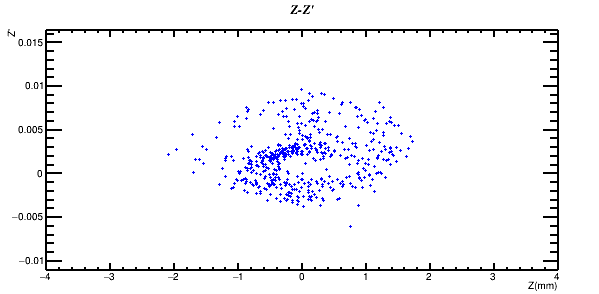}
\par\end{centering}
\caption{$z-z'$ phase space comparison between DEMIRCI and TOUTATIS. }
\label{zzp}
\end{figure}

A similar plot set, shown in Figure \ref{zzp} is also made for the
$z'z$ phase space where $z'$ is defined using the velocity of the
synchronous particle ($Vz_{sync}$) along the $z$ direction:

\begin{equation}
z'=(Vz_{i}-Vz_{sync})/Vz_{sync}
\end{equation}

One can thus see the relation between the $z$ position of the particles
and their velocities showing the details of the bunching process.
Both plots have TOUTATIS results at the top and DEMIRCI results at
the bottom. Although it is hard to read the values from the TOUTATIS
plots, the order of magnitude of the phase spaces coincide and both
programs show similar bunching structures.



\subsection{Acceptance calculation and plots}

\begin{figure}[!htb]
\centering \includegraphics[clip,width=1\columnwidth]{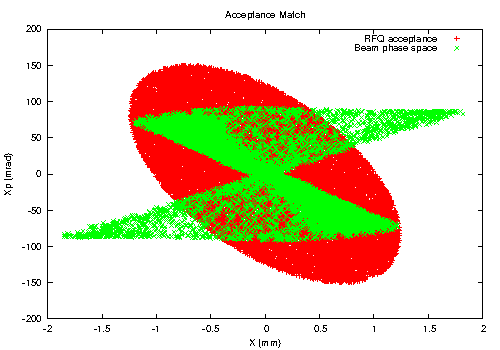} \caption{RFQ beam acceptance and incoming beam's phase space superimposed.
This plot helps in visualising the mismatch factor $M$. }
\label{acceptance} 
\end{figure}

The acceptance of the RFQ under design, is estimated using a very
simple method: a large number of particles are created at the entrance
of the RFQ with a uniform distribution in a large portion of the phase
space. The particles are then moved through the RFQ using the beam
dynamics under the 8T potential as discussed previously. The initial
phase space positions of the particles that survive the trip and exit
the RFQ, constitute the RFQ acceptance as shown in Fig.\ref{acceptance}.
Based on this information the phase space parameters of the RFQ ``\emph{matched}''
acceptance ellipse can be found: $\alpha_{m}$, $\beta_{m}$ and $\gamma_{m}$
. Then the mismatch ($M$) with any incoming beam with known (measured)
parameters can be found as \cite{linacbook}:

\begin{equation}
M=\sqrt{1+\frac{{\Delta+\sqrt{(\Delta+4)\Delta}}}{2}}-1
\end{equation}

where $\Delta=\Delta_{\alpha}^{2}-\Delta_{\beta}\Delta_{\gamma}$
and a delta with subscript represents the difference between the beam's
Twiss parameters and the ``$matched$'' RFQ values, e.g. $\Delta_{\alpha}=\alpha-\alpha_{m}$.

\section{Conclusions and outlook}

The GUI based RFQ design software DEMIRCI is being used and continuously
developed due to needs arising from constant utilisation. It now has
the capability to compute the eight term potential using a 3 dimensional
finite element method. The computed eight term potential can be used
in beam dynamics simulations along with the simplistic two term potential.
It can also guide the user for acceptance matching. The current version
is considered to be mature enough for being used by non-developers as a beta test 
\footnote{DEMIRCI authors can be contacted by email at the
following address: demirci.info@gmail.com }. 
Calculation of the space charge effects
and the study of the construction and assembly errors are next to
be implemented. The design process will also be enriched and automatised
with an extended tool set.


\section{Acknowledgment}

This project has been supported by TUBITAK with project number 114F106 and 117F143. The authors are grateful to Erkcan Ozcan for a careful reading of the manuscript.
\FloatBarrier
\bibliography{Demircirefs.bib}	

\end{document}